\documentstyle[12pt,epsfig]{article}

\newlength{\dinwidth}
\newlength{\dinmargin}
\input{tcilatex}

\begin{document}



\thispagestyle{empty} \vspace*{1cm}

{\Large {\bf Recent Research Developments in Physics}}

\vspace*{1cm}

\begin{center}
{\LARGE {\bf A UNIFYING CONFORMAL FIELD THEORY \\[0pt]
\vspace*{.2cm}  APPROACH TO THE QUANTUM HALL EFFECT} }

\vspace{8mm}

{\large Gerardo Cristofano }
\end{center}

{\large {\footnotesize Dipartimento di Scienze Fisiche,}{\it \ 
{\footnotesize Universit\'{a} di Napoli \textquotedblleft Federico
II\textquotedblright\ \newline
and INFN, Sezione di Napoli}-}{\small Via Cintia - Compl.\ universitario M.
Sant'Angelo - 80126 Napoli, Italy} }

{\large Giuseppe Maiella }

{\large {\footnotesize Dipartimento di Scienze Fisiche,}{\it \ 
{\footnotesize Universit\'{a} di Napoli \textquotedblleft Federico
II\textquotedblright\ \newline
and INFN, Sezione di Napoli}-}{\small Via Cintia - Compl.\ universitario M.
Sant'Angelo - 80126 Napoli, Italy} }

{\large Vincenzo Marotta }

{\large {\footnotesize Dipartimento di Scienze Fisiche,}{\it \ 
{\footnotesize Universit\'{a} di Napoli \textquotedblleft Federico
II\textquotedblright\ \newline
and INFN, Sezione di Napoli}-}{\small Via Cintia - Compl.\ universitario M.
Sant'Angelo - 80126 Napoli, Italy}}

\begin{center}
{\large Adele Naddeo }
\end{center}

{\large {\footnotesize Dipartimento di Scienze Fisiche,}{\it \ 
{\footnotesize Universit\'{a} di Napoli ``Federico II'' \newline
and Coherentia-INFM, Unit\`{a} di Napoli}-}{\small Via Cintia - Compl.\
universitario M. Sant'Angelo - 80126 Napoli, Italy} }

{\large Giuliano Niccoli }

{\large {\it {\footnotesize Sissa and INFN, Sezione di Trieste - Via Beirut
1 - 34100 Trieste, Italy}} }

\begin{center}
\vspace{2cm}

RUNNING TITLE: A UNIFYING CFT FOR THE QUANTUM HALL EFFECT

\pagebreak

{\bf ABSTRACT}
\end{center}

\begin{quotation}
We review the main results of the effective description of the Quantum Hall
fluid for the Jain fillings, $\nu =\frac{m}{2pm+1}$, and the non-standard
ones $\nu =\frac{m}{pm+2}$ by a conformal field theory (CFT) in two
dimensions. It is stressed the unifying character of the $m$-reduction
procedure to construct appropriate twisted CFT models, called Twisted Models
(TM), which by construction reproduce the Quantum Hall topological
properties at those fillings. Indeed for the Jain plateaux we find that the
different descriptions given in the literature fall into different sectors
of the TM for the torus topology. Other interesting aspects are explicitly
seen for the $m=2$ non standard filling $\nu =\frac{1}{p+1}$ (the pairing
case) as the merging of non-Abelian statistics or the instability of the TM
model ($c=2$) versus the Moore-Read one ($c=\frac{3}{2}$). Furthermore by
using Boundary CFT techniques the presence of localized impurities and/or
dissipation is shown to be closely connected with the twisted sector of the
TM, whose presence assures the consistency of the construction and whose
role in describing non trivial global properties of 2D quantum condensed
matter systems is still under study.
\end{quotation}


\pagebreak

\end{document}